\documentclass[aps,prb,twocolumn,showpacs,amsfonts,superscriptaddress]{revtex4-1}
\usepackage{epsfig}\usepackage{subfigure}
\usepackage{graphicx,graphics}
\usepackage{amsmath,ulem}
\usepackage{amssymb}
\usepackage{bm}
\usepackage{color}
\usepackage[utf8]{inputenc}

\usepackage[urlcolor=blue,colorlinks=true,citecolor=blue,linkcolor=blue,pdfstartview={FitH},bookmarks=false]{hyperref}

\begin{document}

\title{Volkov-Pankratov states in topological superconductors}

\author{David J. Alspaugh}
\affiliation{Department of Physics and Astronomy, Louisiana State University, Baton Rouge, LA 70803-4001}

\author{Daniel E. Sheehy}
\affiliation{Department of Physics and Astronomy, Louisiana State University, Baton Rouge, LA 70803-4001}

\author{Mark O. Goerbig}
\affiliation{Laboratoire de Physique des Solides, B\^at. 510, Universit\'e Paris-Saclay, CNRS UMR 8502, F-91405 Orsay Cedex}

\author{Pascal Simon}
\affiliation{Laboratoire de Physique des Solides, B\^at. 510, Universit\'e Paris-Saclay, CNRS UMR 8502, F-91405 Orsay Cedex}

\date{February 12, 2020}

\begin{abstract}

{We study the in-gap states that appear at the boundaries of both 1D and 2D topological superconductors. While the massless Majorana quasiparticles are guaranteed to arise by the bulk-edge correspondence, we find that they could be accompanied by massive Volkov-Pankratov (VP) states which are present only when the interface is sufficiently smooth. These predictions can be tested in an s-wave superconductor with Rashba spin-orbit coupling placed on top of a magnetic domain wall. We calculate the spin-resolved local density of states of the VP states about the band inversion generated by a magnetic domain wall and find that they are oppositely spin-polarized on either side of the topological phase boundary. We also demonstrate that the spatial position, energy-level spacing, and spin polarization of the VP states can be modified by the introduction of in-plane electric fields.}

\end{abstract}

\pacs{}

\maketitle

\section{Introduction}
%The examination of condensed matter systems through the lens of topology has led to the discovery of many new classes of materials.\cite{Hasan2010,Fu2006,Fu2007,Qi2011} Topological superconductors (TSCs), for instance, belong to a distinct quantum phase apart from trivial superconductors (SCs) despite respecting the same global symmetries.~\cite{Schnyder2008} Topological materials can be generically defined by the presence of localized states on their boundaries which are insensitive to perturbations. TSCs in particular have been predicted to host Majorana bound states which are quasiparticles capable of performing quantum computations.~\cite{Fu2008,Read2000}

Topological superconductors (TSCs) are materials which are predicted to host Majorana quasiparticles: excitations which behave as their own antiparticles.~\cite{Majorana1937, Read2000, Beenakker2013, Alicea2012, Qi2011} These quasiparticles obey non-Abelian statistics, making them promising candidates for the topological qubits necessary for fault-tolerant topological quantum computation.~\cite{Kitaev2003, Wilczek2009, Nayak2008} TSCs have been suggested to appear in a variety of condensed matter systems, including strong spin-orbit coupled semiconductor-superconductor (SC) hybrid devices,~\cite{Fu2008, Lutchyn2010} fractional quantum Hall systems at filling factor $\nu = 5/2$,~\cite{Moore1991, Stern2010} spinless $p_{x} + ip_{y}$ SCs,~\cite{Read2000} topological insulator-SC heterostructures,~\cite{Fu2008, Alspaugh2018} integer quantum Hall insulators covered by conventional s-wave SCs,~\cite{Qi2010} and thin films of transition metal dichalcogenides.~\cite{Yuan2014, Hsu2017} 

Majorana zero modes (MZMs) are a 0D version of the Majorana quasiparticle that exist at strictly zero energy and are predicted to emerge at the ends of TSC nanowires and within TSC vortex cores.~\cite{Oreg2010, Read2000, Fu2008} While spectroscopic observations have provided promising signatures for their presence within these systems, it is difficult to energetically resolve the contributions from other effects such as Kondo correlations, Andreev bound states, weak antilocalization, and reflectionless tunneling.~\cite{Mourik2012, Nadj-Perge2014, Das2012, Lee2014, Zitko2015, Sau2015} Recent proposals have instead focused on 1D realizations of Majorana quasiparticles known as chiral Majorana modes (CMMs), which can be found on the boundaries of 2D TSCs. These CMMs are claimed to be responsible for the half-integer quantized conductance plateaus recently observed within quantum anomalous Hall insulator-SC hybrid structures, and have been predicted to be capable of performing quantum computational processes.~\cite{He2017, Chen2017, Lian2018} However, these claims are also under dispute as current research suggests that these half-quantized conductance plateaus can emerge from non-topological sources and are not predicated on the presence of CMMs within the system.~\cite{Kayyalha2020, Huang2018, Ji2018} It is therefore evident that proper identification of MZMs and CMMs requires additional experimental signatures of their emergence.

\begin{figure}
\includegraphics[width=\columnwidth]{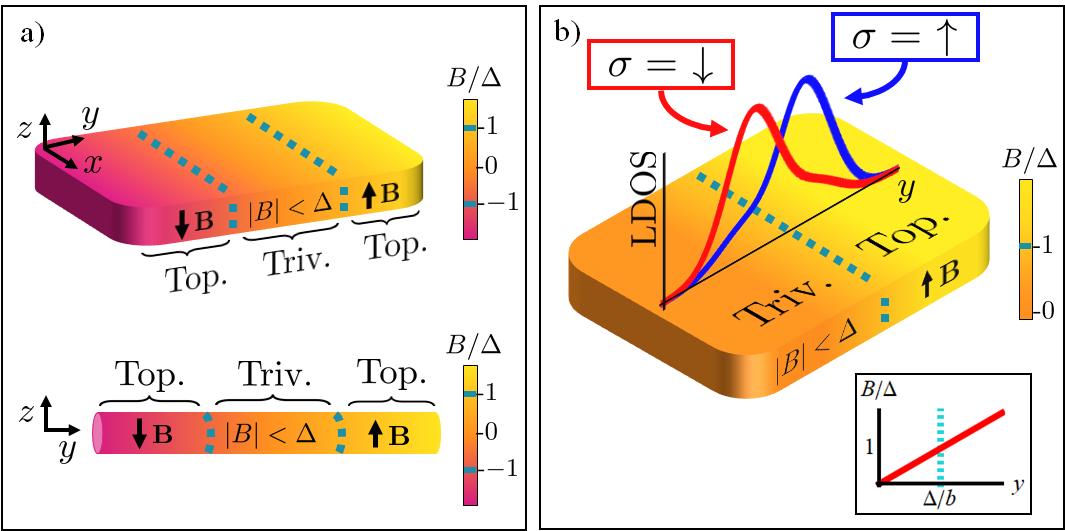}
\caption{(a) Schematic of a magnetic domain wall present in both a two-dimensional (top) and one-dimensional (bottom) SC. Band inversions occur when the energy gap $E_{g} = |B \pm \Delta|$ approaches zero (here $B$ is the Zeeman field and $\Delta$ is the SC gap). These band inversions define TPBs exhibiting Majorana quasiparticles, denoted by the dotted lines. (b) When the transition is sufficiently smooth, the spin-resolved LDOS becomes spin polarized due to the emergence of low-energy VP states, with opposite spin polarizations on either side of the TPB. The inset shows the linear spatial variation of the Zeeman exchange field.}
\label{schem}
\end{figure}

In 1985 Volkov and Pankratov showed that semiconductor junctions with mutually inverted bands can result in the emergence of both massless states and massive states localized at the interface.~\cite{Volkov1985, Karzig2013, Inhofer2017, Mahler2019, Mukherjee2019, Lu2019} If the transition is sharp, only the massless state can be observed below the band gap of each material. In the language of topology, this massless state happens to be the topologically protected edge state whose existence is dictated by the bulk-edge correspondence underlying topological materials.~\cite{Qi2011} However, if the transition is sufficiently smooth massive states may also be observed. Junctions between topological and trivial materials, such as TSCs and SCs, result in similar band inversions which generate the Majorana bound state.~\cite{Menard2017, Rosen2017, Palacio-Morales2019, Li2016, Rontynen2015, Rachel2017, Bjornson2018, Ronetti2019, Kammhuber2017} The purpose of this paper is then to study the properties of VP states which accompany band inversions in SCs with smooth domain walls. Even though these massive states are not intrinsically topological, we stress that they appear as a consequence of the smooth transition between two SCs with different topological nature. While in this article we shall consider only magnetic domain walls, our predictions also apply to smooth transitions resulting from the variation of any other parameter controlling the topological index, such as the chemical potential.

We therefore demonstrate that smoothly varying band inversions generated by magnetic domain walls in both 1D and 2D SCs, schematically shown in Fig.~\ref{schem}~(a), generate massive states in addition to Majorana quasiparticles. In Sec.~\ref{Model} we introduce a minimal model of an s-wave SC with Rashba spin-orbit coupling (SOC), where the band inversion is driven by a magnetic domain wall. We show that zero-energy Majorana excitations are accompanied by massive Volkov-Pankratov (VP) states whose energy gap is determined by the slope of the magnetic Zeeman exchange field associated with the domain wall. Furthermore, while the Majorana excitations are localized about the topological phase boundaries (TPBs) which are denoted by the dotted lines in Fig.~\ref{schem}, we show that the VP states split in real space away from these boundaries. We then analyze how the Majorana and VP states are affected by in-plane electric fields. Despite being electrically neutral, we show that their spatial positions may be controlled through the electric field strength. In addition, the energy level spacing of the VP states decreases as the electric field strength is increased. This is in line with the behavior of VP states in other topological materials, such as topological insulators\cite{Inhofer2017, Tchoumakov2017a} or Weyl semimetals,\cite{Mukherjee2019, Tchoumakov2017} and finds its origin in the relativistic decrease of level spacings due to Lorentz boosts.\cite{Lukose2007} In Sec.~\ref{setups} we calculate the spin-resolved Local Density of States (LDOS) in the vicinity of a TPB and apply these results to both a 1D nanowire and a 2D monolayer. We find that the VP states are spatially spin polarized with opposite polarizations on either side of the TPB, schematically shown in Fig.~\ref{schem}~(b). We predict this will be an observable signature through spin-resolved scanning tunneling spectroscopy measurements. In Sec.~\ref{Conclusion} we summarize our results and discuss the experimentally observable consequences of the VP states.

\section{Model}\label{Model}
In this work we analyze the emergence of VP states in two systems: a 1D nanowire and a 2D monolayer. Let us assume that both of these systems lie in the $x-y$ plane with a Zeeman exchange field $B(\bm{r})$ along the $z$-axis, as shown in Fig.~\ref{schem}. The normal state Hamiltonian for electrons with band mass $m$ in a monolayer system may be given by
\begin{equation}
H_{\text{layer}} = \bigg(-\dfrac{\nabla^{2}}{2m} - \mu\bigg)\sigma_{0} + \alpha(\bm{\sigma}\times -i\bm{\nabla})_{z} + B(\bm{r})\sigma_{z}.
\label{Ham}
\end{equation}
Here and in the remainder of this paper we use a system of units with $\hbar = 1$. The parameter $\alpha$ is the strength of the Rashba SOC, $\mu$ is the chemical potential, and $\sigma_{i}$ are the Pauli matrices in spin space. To model a nanowire, we may remove the degrees of freedom along the $x$-direction in the above Hamiltonian to obtain
\begin{equation}
H_{\text{wire}} = \bigg(-\dfrac{1}{2m}\partial_{y}^{2} - \mu\bigg)\sigma_{0} - i \alpha \partial_{y}\sigma_{x} + B(y)\sigma_{z}.
\end{equation}
In this section we shall explicitly derive the solutions for the monolayer system, and note that the same techniques can equally be applied to $H_{\text{wire}}$. The Bogoliubov-de Gennes (BdG) Hamiltonian is then
\begin{align}
\begin{split}
\mathcal{H} &= \dfrac{1}{2}\int d\bm{r}\Psi^{\dagger}(\bm{r}) H \Psi(\bm{r}),
\\ H &= \bigg(-\dfrac{\nabla^{2}}{2m} - \mu\bigg)\tau_{z}\sigma_{0} + B(\bm{r})\tau_{z}\sigma_{z}
\\ &\hspace{1cm} + \alpha(-i\partial_{y})\tau_{0}\sigma_{x} - \alpha(-i\partial_{x})\tau_{z}\sigma_{y} - \Delta\tau_{y}\sigma_{y}.
\end{split}
\label{BdGHam}
\end{align}
Here $\Psi(\bm{r}) = (\psi_{\uparrow}(\bm{r}),\psi_{\downarrow}(\bm{r}),\psi_{\uparrow}^{\dagger}(\bm{r}),\psi_{\downarrow}^{\dagger}(\bm{r}))^{T}$ is the Nambu spinor with $\psi_{\sigma}(\bm{r})$ being the electron field operators, $\tau_{i}$ are the Pauli matrices for the particle-hole space of the BdG Hamiltonian, and $\Delta$ is the s-wave superconducting order parameter chosen to be real and positive. In the case that the magnetic field is homogeneous, such that $B(\bm{r}) = B$, the energy spectrum of this system has an energy gap at zero momentum given by
\begin{equation}
E_{g} = \left|B \pm \sqrt{\Delta^{2} + \mu^{2}}\right|.
\end{equation}
For positive Zeeman exchange field, we see that the band gap closes at $B = \sqrt{\Delta^{2} + \mu^{2}}$. In the presence of a spatial variation in $B$, there will be a topological phase transition from a Zeeman-dominated region ($B > \sqrt{\Delta^{2} + \mu^{2}}$) to a pairing-dominated region ($B < \sqrt{\Delta^{2} + \mu^{2}}$). Standard arguments based on topology show that the boundary between these regions shall lead to a CMM.\cite{Oreg2010} Here, we show that such a chiral Majorana state can be accompanied by additional massive VP states. To focus on the small momenta about the TPB, we neglect the kinetic energy term ($-\nabla^{2}/2m$) in our Hamiltonian and set $\mu = 0$.~\cite{Kim2015}

We observe that the band inversion is controlled by the Zeeman term $B(\bm{r})$. In order to study a smooth interface between topological and trivial regions, we consider an exchange field profile localized around $y = 0$ and keep only the linear term in its Taylor expansion,~\cite{Oreg2010}
\begin{equation}
B(y) = by.
\label{slope}
\end{equation}
Here the magnetic exchange field has a slope $b>0$ and a characteristic length scale given by $y_{0} = \Delta/b$. To maintain the consistency of our low energy treatment, we shall later ensure that the localization length of the model's bound states is smaller than $y_{0}$. We also note that, as discussed in Ref.~\onlinecite{Tchoumakov2017a}, the linear domain wall profile presented in Eq.~\eqref{slope} yields the same qualitative spectra as other choices of smoothly varying functions, such as $\tanh(y)$.\cite{Lu2019, Tchoumakov2017a}

\subsection{Spectrum of the system}

In the presence of an exchange field described by Eq.~\eqref{slope} our system exhibits two TPBs at $\pm y_{0}$, as shown by the dotted lines in Fig.~\ref{schem}. Our next task is then to find the low-energy Majorana and VP states near these positions. To see this in our model, we transform the $4\times 4$ matrix Hamiltonian of Eq.~\eqref{BdGHam} using
\begin{equation}
U = \dfrac{e^{-i\pi/4}}{2}\begin{pmatrix}
-1 & 1 & -1 & 1
\\ -i & i & i & -i
\\ i & i & -i & -i
\\ 1 & 1 & 1 & 1
\end{pmatrix}.
\end{equation}
Since our system is translationally invariant along the $x$-direction, we write $\psi_{\sigma}(\bm{r}) = \tfrac{1}{\sqrt{L}}\sum_{k}e^{ikx}\psi_{k\sigma}(y)$, where $k$ is the momentum along the $x$-direction and $L$ is the length of the system. These allow us to write our Hamiltonian in a more suggestive form,
\begin{align}
\begin{split}
&\mathcal{H} = \dfrac{1}{2}\sum_{k}\int dy [U^{\dagger}\Psi_{k}(y)]^{\dagger}
\\ &\times \begin{pmatrix}
-\alpha k & 0 & \sqrt{2\alpha b}a_{-}^{\dagger} & 0
\\ 0 & -\alpha k & 0 & \sqrt{2\alpha b}a_{+}^{\dagger}
\\ \sqrt{2\alpha b}a_{-} & 0 & \alpha k & 0
\\ 0 & \sqrt{2\alpha b} a_{+} & 0 & \alpha k
\end{pmatrix} [U^{\dagger}\Psi_{k}(y)].
\end{split}
\label{rotham}
\end{align}
Here $\Psi_{k}(y) = (\psi_{k\uparrow}(y),\psi_{k\downarrow}(y),\psi_{-k\uparrow}^{\dagger}(y),\psi_{-k\downarrow}^{\dagger}(y))^{T}$, and we have defined the ladder operators $a_{\pm} = \sqrt{\tfrac{b}{2\alpha}}[(y \pm y_{0}) + \tfrac{\alpha}{b}\partial_{y}]$. These are harmonic oscillator ladder operators defining states localized at $\mp y_{0}$, respectively. Using the easily obtained eigenvectors of Eq.~\eqref{rotham}, we can then obtain the following eigenvectors of the matrix Hamiltonian in Eq.~\eqref{BdGHam} (within the same approximations and with $-i\partial_{x}\to k$),
\begin{align}
\begin{split}
\varphi_{kn}^{-}(y) &= U A_{kn}\begin{pmatrix}
\phi_{|n|}^{-}(y) \\ 0 \\ Q_{kn}\phi_{|n|-1}^{-}(y) \\ 0
\end{pmatrix},
\\ \varphi_{kn}^{+}(y) &= U A_{kn}\begin{pmatrix}
0 \\ \phi_{|n|}^{+}(y) \\ 0 \\ Q_{kn}\phi_{|n|-1}^{+}(y)
\end{pmatrix}.
\end{split}
\label{varphi}
\end{align}
Here $n$ is an integer, and $\phi_{|n|}^{\pm}(y)$ are the Hermite functions which are eigenfunctions of the operators $a_{\pm}^{\dagger}a_{\pm}$ with eigenvalues $|n|$, given by
\begin{equation}
\phi_{|n|}^{\pm}(y) = \dfrac{\bigg(\dfrac{b}{\pi\alpha}\bigg)^{1/4}}{\sqrt{2^{|n|}|n|!}} e^{-\tfrac{b}{2\alpha}(y \pm y_{0})^{2}} H_{|n|}\bigg[\sqrt{\dfrac{b}{\alpha}}(y \pm y_{0})\bigg].
\end{equation}
Here $H_{|n|}(z)$ are the Hermite polynomials. We can see that these states are localized at $\mp y_{0}$, while the spatial extent of the wavefunctions are determined by the localization length $\ell = \sqrt{\alpha/b}$. The factors $A_{kn}$ and $Q_{kn}$ are given by
\begin{align}
\begin{split}
A_{kn} &= \begin{cases} 1, & n = 0
\\ \dfrac{1}{\sqrt{2}}\sqrt{\dfrac{2\alpha b |n|}{E_{n}(k)^{2} + \alpha k E_{n}(k)}}, & n \neq 0
\end{cases},
\\ Q_{kn} &= \begin{cases} 0, & n = 0
\\ \dfrac{\alpha k + E_{n}(k)}{\sqrt{2\alpha b |n|}}, & n \neq 0
\end{cases}.
\end{split}
\end{align}
The energy eigenvalues of the BdG Hamiltonian are found to be
\begin{equation}
E_{n}(k) = \begin{cases} -\alpha k, & n = 0
\\ \text{sgn}(n)\sqrt{(\alpha k)^{2} + 2\alpha b |n|}, & n \neq 0
\end{cases}.
\label{firstenergies}
\end{equation}
This demonstrates that the CMMs ($n = 0$), localized at $\mp y_{0}$, are generically accompanied by massive VP states ($n\neq 0$). As the transition becomes sharp, the parameter $b$ increases and pushes the energy of the VP states above the superconducting gap $\Delta$. Therefore, the VP states are only observable when the transition from non-magnetic to magnetic regions is sufficiently smooth. From the above expression we see that the first ($n = \pm1$) VP states enter the gap of the SC at the critical slope $b_{c} = \Delta^{2}/2\alpha$, signaling the emergence of VP states into the system. We require that the localization length $\ell$ be smaller than the TPB length scale $y_{0}$, $\ell \ll y_{0}$, which is equivalent to assuming $\sqrt{\alpha b} \ll \Delta$. Indeed, one sees from Eq.~\eqref{firstenergies} that this condition is equivalent to an energy of the first VP states begin below the bulk gap $\Delta$. Otherwise, the VP states would simply not be visible.

To analyze the charge of the CMMs and VP states we calculate the expectation value of the charge operator $\Xi = \text{diag}(e,e,-e,-e)$ with respect to $\varphi_{kn}^{\pm}(y)$, where here $e$ is the electron charge. A quick calculation shows that these expectation values are always zero for both the CMMs and VP states, implying that they are electrically neutral. As discussed in Ref.~\onlinecite{Li2018}, this feature is a consequence of setting $\mu = 0$.

To diagonalize the BdG Hamiltonian we set $\Psi_{k}(y) = \sum_{n}(\varphi_{kn}^{-}(y)\gamma_{kn} + \varphi_{kn}^{+}(y)\beta_{kn})$ in Eq.~\eqref{rotham} and obtain
\begin{equation}
\mathcal{H} = \dfrac{1}{2}\sum_{kn}E_{n}(k)(\gamma_{kn}^{\dagger}\gamma_{kn}^{\phantom{\dagger}} + \beta_{kn}^{\dagger}\beta_{kn}^{\phantom{\dagger}}).
\label{diagBdG}
\end{equation}
Here $\gamma_{kn}$ and $\beta_{kn}$ are the annihilation operators for the bound states localized at $+y_{0}$ and $-y_{0}$ respectively. By expressing $\gamma_{kn}$ and $\beta_{kn}$ in terms of the electron field operators it can be shown that they each obey the relations $\gamma_{kn}^{\dagger} = \gamma_{-k-n}$ and $\beta_{kn}^{\dagger} = -\beta_{-k-n}$. When $n = 0$ these expressions reduce to the Majorana criterion, according to which the Majorana quasiparticle is identical to its own antiparticle.~\cite{Linder2010} We emphasize that Eq.~\eqref{diagBdG} is a low-energy Hamiltonian, and that the summation over $n$ should only include those states that can be observed below the superconducting gap. From the form of $\Psi_{k}(y)$ we can express the electric field operators in terms of the bound state ladder operators,
\begin{align}
\begin{split}
\psi_{k\sigma}(y) &= \sum_{n}\Big(B_{kn\sigma}(y)\gamma_{kn} + C_{kn\sigma}(y)\beta_{kn}\Big),
\\ B_{kn\sigma}(y) &= \begin{cases} \tfrac{e^{i\tfrac{5\pi}{4}}}{2} A_{kn}[\phi_{|n|}^{-}(y) + Q_{kn}\phi_{|n|-1}^{-}(y)], & \sigma = \uparrow
\\ \tfrac{e^{i\tfrac{7\pi}{4}}}{2} A_{kn}[\phi_{|n|}^{-}(y) - Q_{kn}\phi_{|n|-1}^{-}(y)], & \sigma = \downarrow
\end{cases},
\\ C_{kn\sigma}(y) &= \begin{cases} \tfrac{e^{i\tfrac{\pi}{4}}}{2} A_{kn}[\phi_{|n|}^{+}(y) + Q_{kn}\phi_{|n|-1}^{+}(y)], & \sigma = \uparrow
\\ \tfrac{e^{i\tfrac{3\pi}{4}}}{2} A_{kn}[\phi_{|n|}^{+}(y) - Q_{kn}\phi_{|n|-1}^{+}(y)], & \sigma = \downarrow
\end{cases}.
\end{split}
\label{electronoperator}
\end{align}
If we then assume that the TPBs at $\mp y_{0}$ are sufficiently far apart such that $\phi_{|n|}^{\pm}(\pm y_{0}) \approx 0$, in agreement with the condition that $\ell \ll y_{0}$, we may focus on the bound states localized at $+y_{0}$ by writing $\psi_{k\sigma}(y) \approx \sum_{n}B_{kn\sigma}(y)\gamma_{kn}$.

\subsection{Effect of an in-plane electric field}\label{electricfield}
\begin{figure}
\includegraphics[width=6cm]{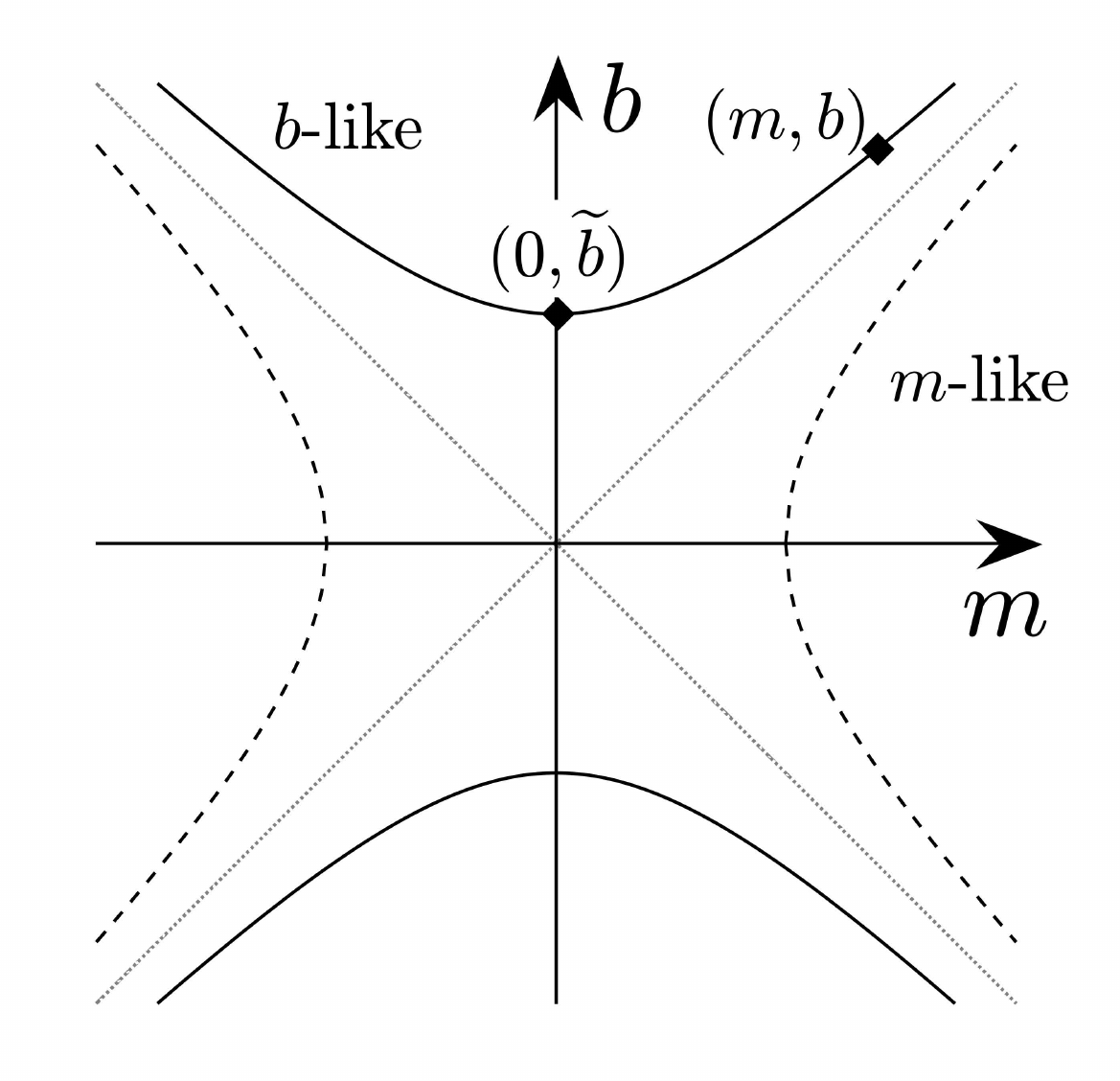}
\caption{Through the use of a hyperbolic transformation, the BdG Hamiltonian of Eq.~\eqref{BdGHamm} can be boosted into a frame in which the electric field vanishes. Whenever $|m| < b$, where $m$ is the electric field strength and $b$ is the slope of the Zeeman field, there exists a Lorentz boost along the black solid line into a frame in which $m = 0$. If instead $|m| > b$, there only exists a Lorentz boost along the dotted lines into a frame in which $b = 0$, removing the TPB and hence the CMM and VP states.}
\label{boost}
\end{figure}
Next, we study the response of the CMMs and VP states under an in-plane electric field along the $y$-direction. To include this in our model, we introduce a spatially varying chemical potential into our Hamiltonian. The electric field is then given by the negative of the gradient of this potential. Still neglecting the kinetic energy term, the normal state Hamiltonian is then
\begin{equation}
H_{\text{layer}} =  - \mu(y)\sigma_{0} + \alpha(\bm{\sigma}\times -i\bm{\nabla})_{z} + B(y)\sigma_{z}.
\end{equation}
Here we define
\begin{equation}
\mu(y) = m y,
\end{equation}
where $m$ is the strength of the electric field which can either be positive or negative. The BdG Hamiltonian may then be written as
\begin{align}
\begin{split}
\mathcal{H} &= \dfrac{1}{2}\sum_{k}\int dy \Psi_{k}^{\dagger}(y)H\Psi_{k}(y),
\\ H &= -\mu(y)\tau_{z}\sigma_{0} + B(y)\tau_{z}\sigma_{z} 
\\ &\hspace{1cm} + \alpha(-i\partial_{y})\tau_{0}\sigma_{x} -\alpha k \tau_{z}\sigma_{y} - \Delta\tau_{y}\sigma_{y}.
\end{split}
\label{BdGHamm}
\end{align}
As shown below, it is sufficient to consider the case $|m| < b$. In this case we shall show that the BdG Hamiltonian may be transformed to a similar form as that of the previous section. To find the eigenvectors of the first quantized Hamiltonian such that $H\varphi = E\varphi$, we apply a hyperbolic transformation that is akin to a Lorentz boost generated by $\exp(\eta\tau_{0}\sigma_{z}/2)$, where $\eta = \tanh^{-1}(m/b)$.~\cite{Tchoumakov2017,Tchoumakov2017a} To perform this, we may rearrange the Schr\"{o}dinger equation by writing 
\begin{equation}
e^{\tfrac{\eta}{2}\tau_{0}\sigma_{z}}(H - E)e^{\tfrac{\eta}{2}\tau_{0}\sigma_{z}}(\mathcal{N}e^{-\tfrac{\eta}{2}\tau_{0}\sigma_{z}}\varphi) = 0,
\end{equation}
where here $\mathcal{N}$ is a normalization constant included because the exponential matrices modify the norm of the wavefunction. By rewriting the exponential matrices we may obtain the Lorentz-boosted Schr\"{o}dinger equation,
\begin{equation}
\widetilde{H}\widetilde{\varphi} = \gamma E \widetilde{\varphi}.
\end{equation}
Here $\gamma = 1/\sqrt{1 - (m/b)^{2}}$ and $\widetilde{\varphi} = \mathcal{N}e^{-\eta\tau_{0}\sigma_{z}/2}\varphi$. We have also defined the Lorentz-boosted BdG Hamiltonian:
\begin{align}
\begin{split}
\widetilde{H} &= \widetilde{b} y \tau_{z}\sigma_{z} - \widetilde{E}\tau_{0}\sigma_{z} 
\\ &\hspace{1cm} + \alpha(-i\partial_{y})\tau_{0}\sigma_{x} -\alpha k \tau_{z}\sigma_{y} - \Delta\tau_{y}\sigma_{y},
\end{split}
\label{boostedBdGHam}
\end{align}
where $\widetilde{b} =b/\gamma$ and $\widetilde{E} = \gamma\tfrac{m}{b}E$. Importantly, the spatially varying terms associated with $\mu(y)$ are no longer present in $\widetilde{H}$. This Hamiltonian is then similar in form to the original BdG Hamiltonian in Eq.~\eqref{BdGHam} of the previous section, with an additional constant term which depends on the energy. From the above expressions, we see that the behavior of this hyperbolic transformation is analogous to that of a Lorentz boost which transforms the electric field into a renormalized magnetic field, as shown in Fig.~\ref{boost}. 

Recall that we assumed that $|m| < b$. If we instead had that $|m| > b$, we would be in the $m$-like quadrant of Fig.~\ref{boost}, and therefore would only be able to boost along the dashed line to a point with zero magnetic field ($b = 0$). Without any Zeeman exchange field, there will be no TPBs and therefore no CMMs nor VP states emerging within the system. This shows that if the electric field is too strong, the bound states are destroyed despite the fact that they are initially electrically neutral. In the following analysis we shall then assume that the slope of the chemical potential is sufficiently smooth such that $|m| < b$.

The similarity of Eq.~\eqref{boostedBdGHam} to Eq.~\eqref{BdGHam} implies that we can find the low-energy CMMs and VP states via a similar approach. We then introduce the following energy-dependent unitary transformation
\begin{align}
\begin{split}
W_{E} &= \dfrac{1}{2}\begin{pmatrix}
-F_{E}^{+} & F_{E}^{-} & -F_{E}^{+} & F_{E}^{-}
\\ -i F_{E}^{+} & i F_{E}^{-} & iF_{E}^{+} & -iF_{E}^{-}
\\ i G_{E}^{+} & i G_{E}^{-} & -iG_{E}^{+} & -iG_{E}^{-}
\\  G_{E}^{+} &  G_{E}^{-} & G_{E}^{+} & G_{E}^{-}
\end{pmatrix},
\\ F_{E}^{\pm} &= \dfrac{\pm\widetilde{E} + \sqrt{\widetilde{E}^{2} + \Delta^{2}}}{\sqrt{\widetilde{E}^{2} + \Delta^{2} \pm \widetilde{E}\sqrt{\widetilde{E}^{2} + \Delta^{2}}}},
\\ G_{E}^{\pm} &= \dfrac{\Delta}{\sqrt{\widetilde{E}^{2} + \Delta^{2} \pm \widetilde{E}\sqrt{\widetilde{E}^{2} + \Delta^{2}}}}.
\end{split}
\end{align}
This allows us to write our Lorentz boosted BdG Hamiltonian in a more suggestive form,
\begin{align}
\begin{split}
W_{E}^{\dagger} &  \widetilde{H}W_{E}\! =\! \begin{pmatrix}
-\alpha k & 0 & \sqrt{2\alpha \widetilde{b}}\widetilde{a}_{-}^{\dagger} & 0
\\ 0 & -\alpha k & 0 & \sqrt{2\alpha \widetilde{b}}\widetilde{a}_{+}^{\dagger}
\\ \sqrt{2\alpha \widetilde{b}}\widetilde{a}_{-} & 0 & \alpha k & 0
\\ 0 & \sqrt{2\alpha \widetilde{b}} \widetilde{a}_{+} & 0 & \alpha k
\end{pmatrix}\! .
\end{split}
\label{rothamm}
\end{align}
Here we have defined the ladder operators $\widetilde{a}_{\pm} = \sqrt{\tfrac{\widetilde{b}}{2\alpha}}[(y \pm \tfrac{\sqrt{\widetilde{E}^{2} + \Delta^{2}}}{\widetilde{b}}) + \tfrac{\alpha}{\widetilde{b}}\partial_{y}]$. The similarity of Eq.~\eqref{rothamm} to Eq.~\eqref{rotham} of the previous section allows us to quickly find the energy eigenvalues of the original Schr\"{o}dinger equation,
\begin{align}
E_{n}(k) &= \begin{cases} -\dfrac{1}{\gamma}\alpha k, & n = 0
\\ \dfrac{1}{\gamma}\text{sgn}(n)\sqrt{(\alpha k)^{2} + 2\alpha\widetilde{b}|n|}, & n\neq 0
\end{cases}.
\label{energies}
\end{align}
The eigenvectors of $H$ in Eq.~\eqref{BdGHamm} are then found to be
\begin{align}
\begin{split}
\varphi_{kn}^{-}(y) &= \dfrac{e^{\eta\tau_{0}\sigma_{z}/2}}{\mathcal{N}_{kn}^{-}} W_{E_{n}(k)} \widetilde{A}_{kn}\begin{pmatrix}
\widetilde{\phi}_{|n|}^{-}(y,k) \\ 0 \\ \widetilde{Q}_{kn}\widetilde{\phi}_{|n|-1}^{-}(y,k) \\ 0
\end{pmatrix},
\\ \varphi_{kn}^{+}(y) &= \dfrac{e^{\eta\tau_{0}\sigma_{z}/2}}{\mathcal{N}_{kn}^{+}} W_{E_{n}(k)} \widetilde{A}_{kn}\begin{pmatrix}
0 \\ \widetilde{\phi}_{|n|}^{+}(y,k) \\ 0 \\ \widetilde{Q}_{kn}\widetilde{\phi}_{|n|-1}^{+}(y,k)
\end{pmatrix}.
\end{split}
\label{mvarphi}
\end{align}
Here, the Hermite functions $\widetilde{\phi}_{|n|}^{\pm}(y,k)$ are once again the eigenfunctions of $\widetilde{a}_{\pm}^{\dagger}\widetilde{a}_{\pm}$ with eigenvalue $|n|$, and are given by
\begin{align}
\begin{split}
\widetilde{\phi}_{|n|}^{\pm}(y,k) &= \dfrac{\bigg(\dfrac{\widetilde{b}}{\pi\alpha}\bigg)^{1/4}}{\sqrt{2^{|n|}|n|!}}  e^{-\tfrac{\widetilde{b}}{2\alpha}(y \pm \tfrac{\sqrt{\widetilde{E}_{n}(k)^{2} + \Delta^{2}}}{\widetilde{b}})^{2}} 
\\ &\hspace{0.5cm} \times H_{|n|}\bigg[\sqrt{\dfrac{\widetilde{b}}{\alpha}}\bigg(y \pm \dfrac{\sqrt{\widetilde{E}_{n}(k)^{2} + \Delta^{2}}}{\widetilde{b}}\bigg)\bigg].
\end{split}
\end{align}
The spatial location of the wavefunctions depends on the energy and electric field strength through $\widetilde{E}_{n}(k) = \gamma\tfrac{m}{b}E_{n}(k)$, while their spatial extent now depends on the localization length $\ell = \sqrt{\alpha/\widetilde{b}}$ which increases with the electric field strength. The terms $\widetilde{A}_{kn}$ and $\widetilde{Q}_{kn}$ are given by
\begin{align}
\begin{split}
\widetilde{A}_{kn} &= \begin{cases} 1, & n = 0
\\ \dfrac{1}{\sqrt{2}}\sqrt{\dfrac{2\alpha \widetilde{b} |n|}{[\gamma E_{n}(k)]^{2} + \alpha k \gamma E_{n}(k)}}, & n \neq 0
\end{cases},
\\ \widetilde{Q}_{kn} &= \begin{cases} 0, & n = 0
\\ \dfrac{\alpha k + \gamma E_{n}(k)}{\sqrt{2\alpha \widetilde{b} |n|}}, & n \neq 0
\end{cases}.
\end{split}
\end{align}
From Eq.~\eqref{energies} we notice that the CMM, given by $n = 0$, still remains at zero energy despite the modification of its wavefunction by the electric field. We may observe that the effect of the electric field, regardless of its direction, is to shift the location of the bound states which are now localized at $y = \mp\sqrt{\widetilde{E}_{n}(k)^{2} + \Delta^{2}}/\widetilde{b}$. This amounts to spatially pushing apart the two TPBs. As shown in Fig.~\ref{schem}~(a) the region between the two TPBs, denoted by the dotted lines, is topologically trivial. When $|m| = b$, the entire system is then covered by the topologically trivial domain. 

Using the wavefunctions in Eq.~\eqref{mvarphi} we may similarly diagonalize the BdG Hamiltonian as in the previous section. However, in this case the $B_{kn\sigma}(y)$ and $C_{kn\sigma}(y)$ coefficients along with the $\mathcal{N}_{kn}^{\pm}$ normalizations no longer have closed form analytic solutions, and must be evaluated numerically. In addition, from Eq.~\eqref{mvarphi} we may numerically calculate that the wavefunctions are no longer electrically neutral at $m \neq 0$.

\section{Experimental consequences}\label{setups}
\subsection{Local density of states}\label{LDOS}
We have shown that the CMMs occuring in TPBs can be generically accompanied by low-energy VP states. Our next task is then to determine signatures of the VP states in the LDOS that is measurable via tunneling spectroscopy. We may determine the LDOS from the spectral function of the system. In the following we focus on the states localized at $+y_{0}$ under the assumption that $\ell < y_{0}$. From the Fourier transform of the electronic retarded Green's function $G^{R}(\bm{r}\sigma t,\bm{r}'\sigma't') = -i\theta(t - t')\langle\{\psi_{\sigma}(\bm{r},t),\psi_{\sigma'}^{\dagger}(\bm{r}',t')\}\rangle$, with $\theta(t)$ the Heaviside step function, we find the spectral function $A(\bm{r}\sigma;\omega) = -2\text{Im} G^{R}(\bm{r}\sigma,\bm{r}\sigma;\omega)$ and thus the spin resolved LDOS as
\begin{equation}
\rho_{\sigma}(\bm{r},\omega) = \dfrac{A(\bm{r}\sigma;\omega)}{2\pi} = \dfrac{1}{L}\sum_{kn}|B_{kn\sigma}(y)|^{2}\delta(\omega - E_{n}(k)).
\label{rho}
\end{equation}
The total LDOS $\rho(\bm{r},\omega)$ is then given as the summation of both spin components. We may then analyze both the 1D nanowire and 2D monolayer systems originally shown in Fig.~\ref{schem}.

\subsection{1D nanowire}
\begin{figure}
\includegraphics[scale=0.7]{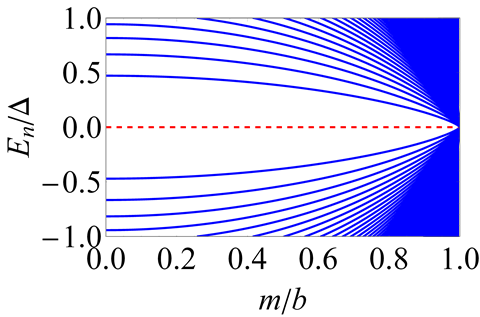}
\caption{Evolution of the nanowire bound state energies as the electric field strength is increased. Here $m$ is the strength of the electric field and $b$ is the slope of the Zeeman field. The MZM is denoted by the red dashed line, while the VP states are denoted by the blue solid lines. While the MZM remains at zero energy, the energy level spacing of the VP states decreases with increased electric field strength relative to the magnetic field slope.}
\label{nanowiredisp}
\end{figure}
\begin{figure}
\includegraphics[width=\columnwidth]{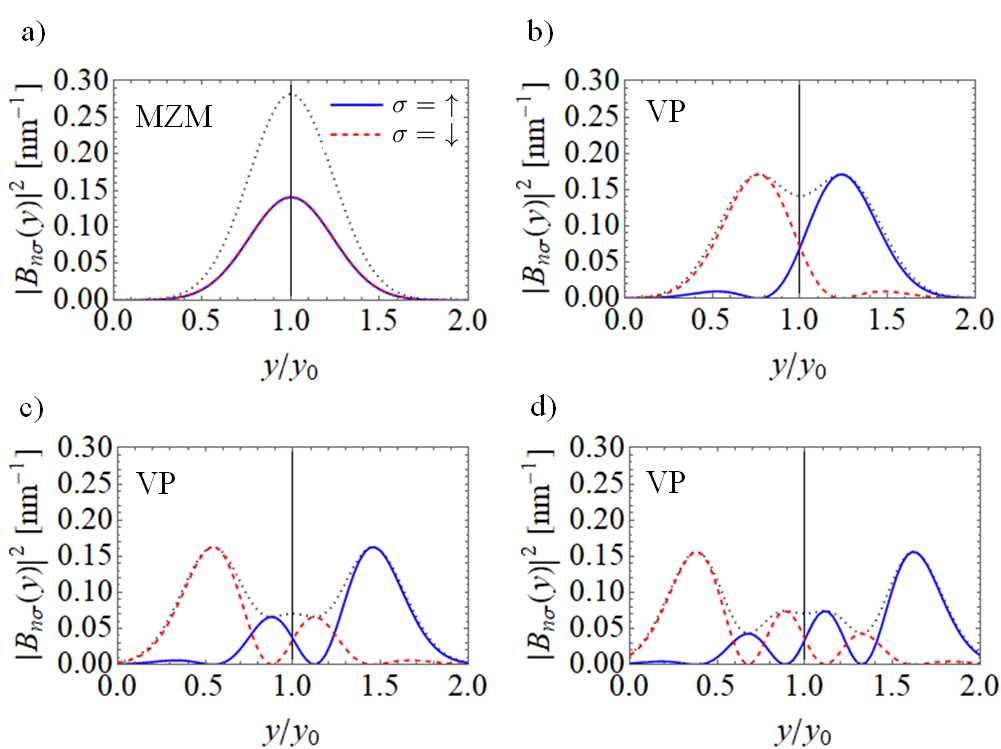}
\caption{Magnitude of the $B_{n\sigma}(y)$ coefficients which appear in the LDOS for the $n = 0,1,2,3$ nanowire bound states in the absence of an electric field, shown in figures (a), (b), (c), and (d), respectively. The vertical black lines mark the TPB at $y_{0} = 3$ nm. The dotted black lines denote the sum of both spin components. The MZM ($n = 0$) state is localized at $y_{0}$, while the VP ($n\neq 0$) states split away from the TPB, with opposite spin polarizations on either side.}
\label{nanowireB}
\end{figure}
\begin{figure}
\includegraphics[width=\columnwidth]{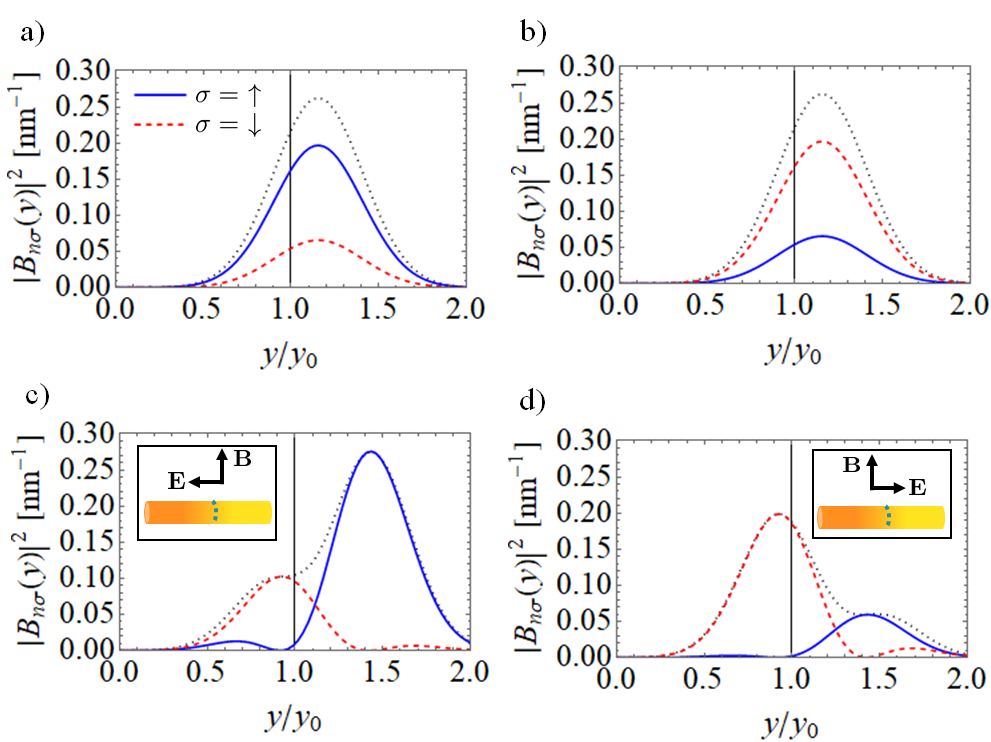}
\caption{Magnitude of the $B_{n\sigma}(y)$ coefficients in the presence of an electric field. Figures (a) and (b) show the MZM ($n = 0$) state as $m = b/2$ and $m = -b/2$ respectively, where $m$ is the strength of the electric field and $b$ is the slope of the Zeeman field. Figures (c) and (d) show the $n = 1$ VP state as $m = b/2$ and $m = -b/2$ respectively. The vertical black line marks $y_{0} =  3$ nm, which is the topological phase boundary as $m = 0$. The insets in (c) and (d) show the orientation of the electric and magnetic fields as $m$ is positive and negative, respectively.}  
\label{nanowireBm}
\end{figure}

Applying the techniques of the previous section to the case of a 1D nanowire, we find a discrete energy spectrum given by $E_{n} = \tfrac{\text{sgn}(n)}{\gamma}\sqrt{2\alpha\widetilde{b}|n|}$, which describes a set of MZMs and additional VP states localized at $y = \pm \sqrt{\widetilde{E}_{n}^{2} + \Delta^{2}}/\widetilde{b}$. Motivated by recent experiments, we set $\Delta = 0.3$ meV, $\alpha = 0.1$ meV~nm, and $b = 0.1$ meV~nm$^{-1}$.~\cite{Menard2017} The superconducting coherence length of the system is then $\xi = \alpha/\Delta\approx 0.33$ nm. In this case we find that $y_{0} = 3$ nm and $\ell = 1$ nm, which maintains the consistency of the low-energy treatment introduced in Eq.~\eqref{slope}. 
%With this choice of parameters the localization length of the Majorana state and the slope of the Zeeman exchange field are consistent with the experimental findings in Ref.~\onlinecite{Menard2017}.

In Fig.~\ref{nanowiredisp} we plot the energy spectrum of the nanowire bound states as a function of the electric field strength. We observe that as the electric field increases, there is a decrease in the energy level spacing of the MZM and VP states, and that more VP states emerge below the superconducting gap.

As $\ell < y_{0}$, we find that the two sets of bound states localized at $\mp y_{0}$ do not overlap in space, and may focus our analysis on the $+y_{0}$ states. The LDOS of the bound states centered at $y_{0}$ is given by $\rho(y,\omega) = \sum_{\sigma n}|B_{n\sigma}(y)|^{2}\delta(\omega - E_{n})$, where here $B_{n\sigma}(y)$ is found by setting $k = 0$ in Eq.~\eqref{electronoperator}. In Fig.~\ref{nanowireB} we plot $|B_{n\sigma}(y)|^{2}$ for the MZM and first three positive VP states for $m = 0$. While the MZM is centered at the TPB, the VP states begin to split in space away from the phase boundary. In addition, while the MZM is not spin polarized, we see that the VP states are strongly spin polarized in space, with opposite spin polarizations on either side of the TPB. We find that these spin polarizations are interchanged for the $n < 0$ states.

In Fig.~\ref{nanowireBm} we demonstrate the effect of the electric field on the bound states by numerically calculating the $B_{n\sigma}(y)$ coefficients as discussed in Sec.~\ref{electricfield}. Regardless of the electric field direction, all of the states are shifted spatially to the right. This shift is easily seen through the TPB location $y = \sqrt{\widetilde{E}_{n}^{2} + \Delta^{2}}/\widetilde{b}$, which acquires an energy-dependent displacement as a consequence of the Lorentz boost. In addition, the localization length increases with increasing field strength, causing the states to spread out in space. At $|m| = b$, we see that all of the states vanish as the localization length diverges. The spin polarization, however, depends on both the strength and direction of the electric field and are no longer equal and opposite in space.

\subsection{2D monolayer}

\begin{figure}
\includegraphics[width=\columnwidth]{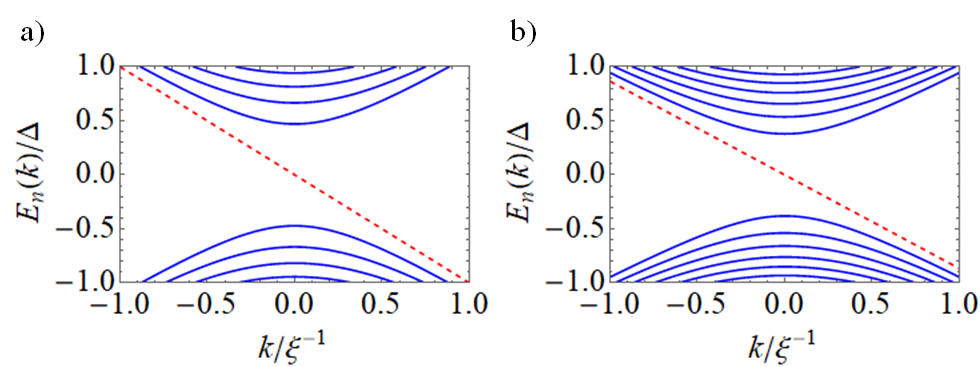}
\caption{Energy spectrum of the monolayer bound states, where $k$ is the momentum along the $x$-direction and $\xi$ is the superconducting coherence length. The electric field is given by $m = 0$ and $m = b/2$ in figures (a) and (b) respectively, where $m$ is the strength of the electric field and $b$ is the slope of the Zeeman field. More VP states (blue solid lines) are introduced as the electric field strength increases, and the slope of the CMM is renormalized (red dashed lines).}
\label{dispersion}
\end{figure}

\begin{figure}
\includegraphics[width=\columnwidth]{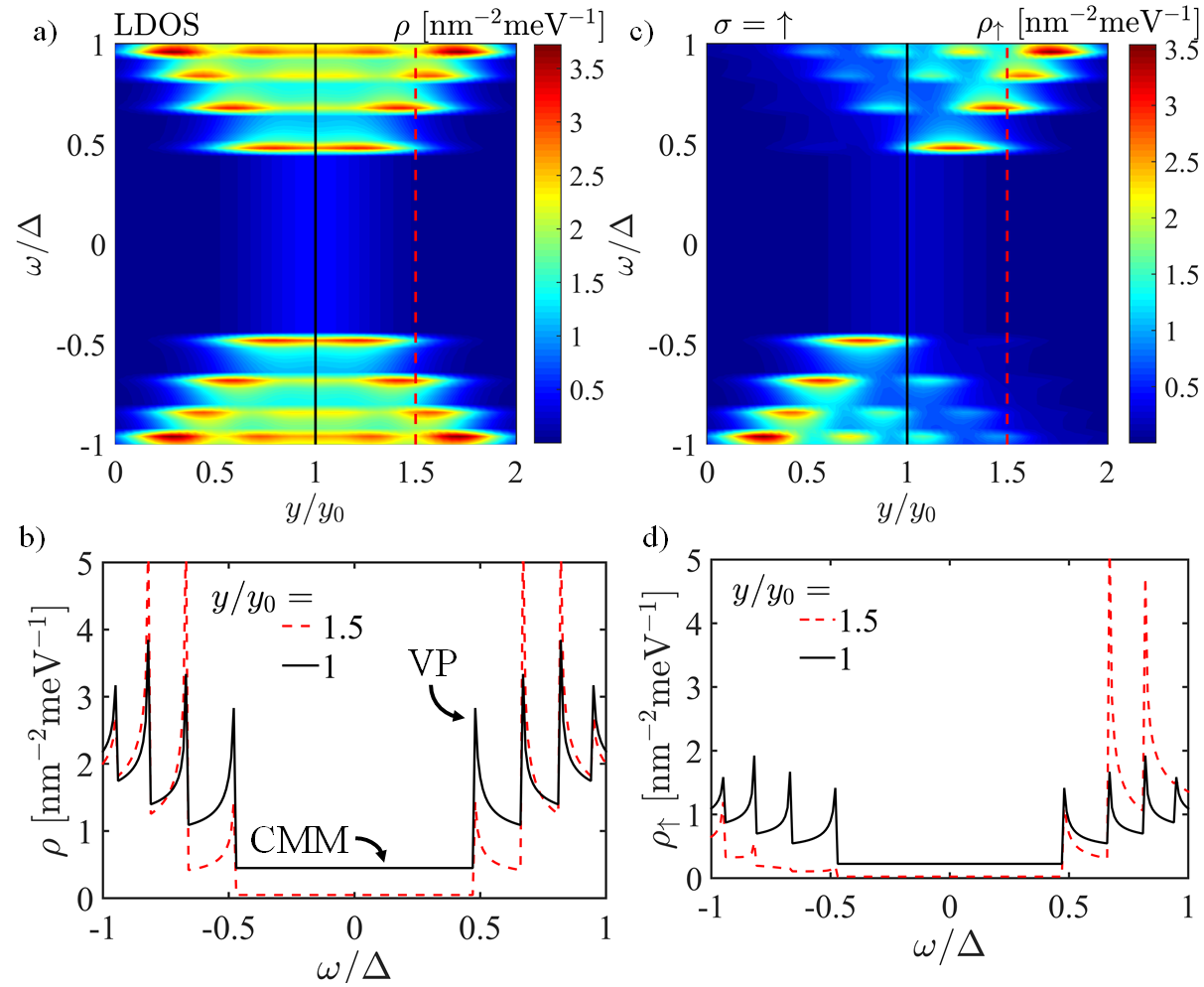}
\caption{(a) Total LDOS of the monolayer system in the absence of an electric field, plotted as a function of position and energy. Here $\Delta = 0.3$~meV is the SC gap, and $y_{0} = 3$~nm is the distance of the CMM to the center of the magnetic domain wall. The solid black and dashed red lines of constant position at $y_{0}$ and $1.5y_{0}$ are individually plotted in Figure (b). The LDOS is constant at low energies due to the linear dispersion of the CMM, while the peaks arise from the VP states. (c) Spin-up component of the spin-resolved LDOS. Lines of constant position are similarly plotted in Figure (d).}
\label{LDOSgraph}
\end{figure}
 
\begin{figure}
\includegraphics[width=\columnwidth]{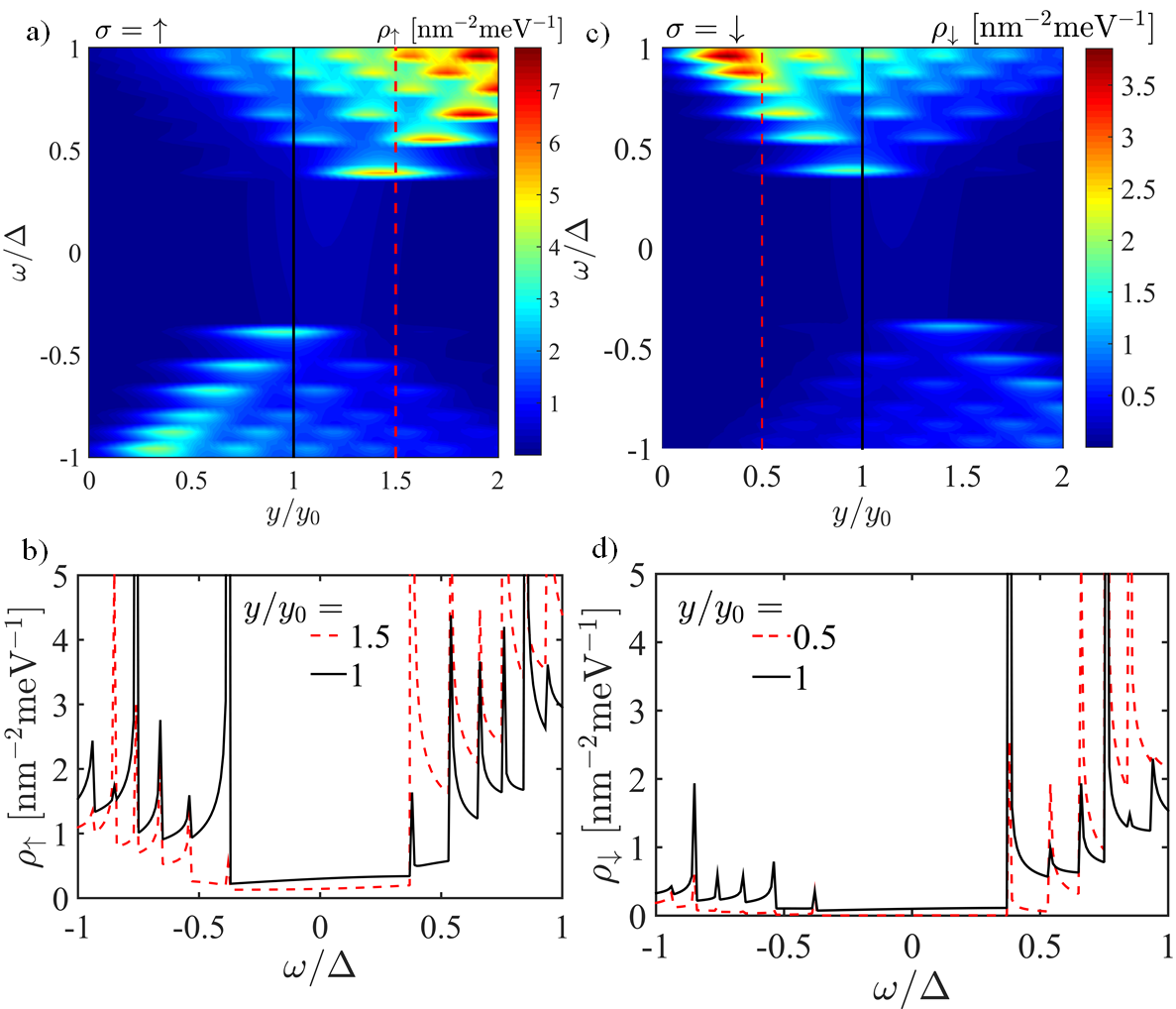}
\caption{(a) Spin-up component of the spin-resolved LDOS of the monolayer system in the presence of an electric field given by $m = b/2$, where $m$ is the strength of the electric field and $b$ is the slope of the Zeeman field. Here $\Delta = 0.3$~meV is the SC gap, and $y_{0} = 3$~nm is the distance of the CMM to the center of the magnetic domain wall in the absence of an electric field. The solid black and dashed red lines of constant position at $y_{0}$ and $1.5y_{0}$ are individually plotted in Figure (b). (c) Spin-down component of the spin-resolved LDOS. Lines of constant position at $y_{0}$ nm and $0.5y_{0}$ nm are individually plotted in Figure (d). The spin-up component is amplified, while the spin-down component is suppressed.}
\label{LDOSgraphm}
\end{figure}

To analyze the case of a 2D monolayer, we adopt the parameter values of the previous section and plot the dispersion of Eq.~\eqref{energies} in Fig.~\ref{dispersion}. In contrast to the nanowire, this system exhibits a CMM which linearly disperses only in one direction parallel to the $x$ axis. Similar to the nanowire case, however, more VP states enter the gap as the electric field strength increases.

To evaluate the LDOS in Eq.~\eqref{rho} we convert the sum over momentum into an integral which we solve analytically and plot in Fig.~\ref{LDOSgraph} as a function of position and energy. Here the LDOS creates an ``X-shape" as a 2D manifestation of what was observed in Fig.~\ref{nanowireB} of the nanowire. For low energies, we see that the spectrum is constant due to the linear dispersion of the CMM localized at $y_{0}$. However, as the energy approaches the beginning of the VP bands, we see that the LDOS has large peaks which split in space about the TPB. 

We note that the ``X-shape" of these peaks in Fig.~\ref{LDOSgraph}~(a) resembles what is experimentally observed in Ref.~\onlinecite{Menard2017}, in which magnetic Co-Si islands are deposited beneath a superconducting Pb monolayer. These Co-Si islands create a spatially varying Zeeman exchange field below the 2D SC, leading to a TPB on their circular edge which hosts CMMs. However, the experimental data unexpectedly displayed additional states apart from the CMM, splitting in space away from the Co-Si edge with nonzero energies. The LDOS above the Co-Si island is plotted in Fig.~2~(g) of Ref.~\onlinecite{Menard2017} as a function of position and energy, and these additional states create an ``X-shape" that is similar to what is analytically derived in Fig.~\ref{LDOSgraph}~(a). We propose that the additional states observed in Ref.~\onlinecite{Menard2017} are in fact VP states which emerge due to the smoothly varying exchange field decaying away from the Co-Si island.

To prove that the additional states found in Ref.~\onlinecite{Menard2017} are truly VP states, we may analyze the spin-resolved LDOS in Fig.~\ref{LDOSgraph}~(c). We find that the spin-up components are largely grouped to the right (left) of the TPB for positive (negative) energy values. The spin-down components are opposite, and are grouped on the left (right) of the TPB for positive (negative) energies. This shows that the spin-up and spin-down components of the VP states shift away from each other in space. These spin-polarizations can be measured via spin-resolved scanning tunneling spectroscopy experiments, and would provide additional evidence that the additional states observed in Ref.~\onlinecite{Menard2017} are indeed VP states.~\cite{Jeon2017} In our treatment we find that the CMM has equal spin-up and spin-down components, and is therefore not spin-polarized. This is a consequence of our $\mu = 0$ assumption, and in general the CMM may exhibit a net spin-polarization for nonzero values of the chemical potential.~\cite{Li2018} In contrast, we expect that the strong spin-polarizations of the VP states will be insensitive to small variations of the chemical potential, leading to easily identifiable signatures in spin-resolved scanning tunneling spectroscopy experiments.

In order to study the LDOS under the presence of an electric field, we once again convert the momentum summation of Eq.~\eqref{rho} into an integral which we evaluate analytically, and then numerically calculate the $B_{kn\sigma}(y)$ coefficients as discussed in Sec.~\ref{electricfield}. In Fig.~\ref{LDOSgraphm} we plot the spin-up and spin-down components of the LDOS once again as functions of position and energy. Similar to the previous section we find that for $m > 0$ the spin-up component is significantly amplified for positive energies, while the spin-down component is suppressed for negative energies. In the case that $m < 0$, however, we find that the spin-down component is amplified for negative energies, while the spin-up component is suppressed for positive energies. In addition, we see that all the bound states move to the right for all nonzero energy values, and that many more peaks appear in the LDOS due to the presence of additional VP states.

\section{Conclusion}\label{Conclusion}
%We have demonstrated that the smooth transition between trivial and topological regions of a TSC-SC interface driven by a magnetic domain wall can generate massive VP states in addition to the massless topologically protected Majorana mode in both 1D and 2D SCs. The energy gap of these bound states may be controlled by the rate of change of the domain wall's exchange field. In order for these bound states to be observed under the superconducting gap, the magnetic domain wall must be sufficiently smooth. We have calculated the spin-resolved LDOS about one of the TPBs, and found that in contrast to the unpolarized Majorana state, the VP states are strongly spin polarized with opposite spin polarizations on either side of the boundary. The total LDOS of the VP states splits in space about the TPB, similar to what is seen in experiment.~\cite{Menard2017} We also find that the VP states are sensitive to in-plane electric fields, which can decrease their energy level spacing and potentially bring them below the superconducting gap. This may allow previously unobservable states to become visible through scanning tunneling spectroscopy experiments. We also find that the locations of both the Majorana and VP states can be manipulated through the strength of an in-plane electric field. For strong enough electric fields, the TSC region is destroyed as the entire system enters the trivial regime. Despite the fact that the Majorana and VP states are electrically neutral, we find that their emergence and spatial positions can be electrically controlled.

In this work we analyzed smooth magnetic domain walls that generate band inversions within 1D and 2D SCs. It is known that band inversions in 1D superconducting nanowires exhibit MZMs fixed at zero energy, while band inversions in 2D superconducting monolayers lead to linearly dispersing CMMs. While modern proposals have focused on CMMs as a potentially more robust alternative to the MZM for topological quantum computation, both MZMs and CMMs have been difficult to experimentally verify. We have shown that if the transition between topological and trivial regions is smooth enough, the massless Majorana quasiparticles are accompanied by massive VP states. These VP states arise purely as a consequence of the transition between topologically different phases. While in this work we have focused on magnetic domain walls, we emphasize that our predictions equally apply to any other smooth transition that results from the variation of a parameter controlling the topological phase.

We have shown that the energy level spacing of the VP states is controlled by the slope of the Zeeman exchange field, and that the VP states are only observable below the superconducting gap when the slope of the exchange field is smaller than a critical value. We also found that while the Majorana states are localized at the TPB, the VP states split in space away from the TPB. In the case of a 2D monolayer, the splitting of the VP states creates an ``X-shape" in the LDOS as a function of position. This X-shape is similar to what is experimentally observed in 2D SCs around spatially extended magnetic Co clusters, and we predict that VP states may be present within these systems.\cite{Menard2017} We calculated how the VP states respond to in-plane electric fields, and have shown that their energy-level spacing depends on the electric field strength. As the strength of the electric field is increased, more VP states are observable below the superconducting gap. We also found that the spatial location of both the Majorana and VP states is controlled by the magnitude of the electric field. If the electric field becomes too strong, both the Majorana and VP states are destroyed as the entire system enters the trivial regime.

We derived the spin-resolved LDOS of the Majorana and VP states in the vicinity of a TPB. In contrast to the Majorana quasiparticles, we have found that the VP states are strongly spin polarized. As the VP states split in space away from the TPB, we have shown that opposite sides of the TPB display opposite spin polarizations. The magnitude of these spin polarizations is dependent on both the strength and direction of the electric field. In the case of zero field, the magnitudes of the spin polarization are equal and opposite on either side of the TPB, while nonzero electric fields lead to an asymmetry of these magnitudes across the TPB. We predict that this will be an observable signature of the VP states via spin-resolved scanning tunneling spectroscopy.

\section{Acknowledgments}
This work has been supported by the Chateaubriand Fellowship Program. Part of  this work was performed at the Aspen Center for Physics, which is supported by National Science Foundation grant PHY-1607611. The authors acknowledge useful discussions with Dirk Morr and Xin Lu. DJA also acknowledges the hospitality of the Laboratoire de Physique des Solides during the Fall of 2018 when part of this work was completed.

\bibliography{biblio} 

\bibliographystyle{apsrev4-1}

\end{document}